\documentclass[12pt]{article}
\begin{document}
\title{{\bf Spacetime Average Density\\(SAD)\\Cosmological Measures}
\thanks{Alberta-Thy-10-14, arXiv:1406.0504 [hep-th]}}

\author{Don N. Page
\thanks{Internet address:
profdonpage@gmail.com}
\\
Department of Physics\\
4-183 CCIS\\
University of Alberta\\
Edmonton, Alberta T6G 2E1\\
Canada}

\date{2014 October 22}
\maketitle
\large
\begin{abstract}
\baselineskip 20.45 pt

The measure problem of cosmology is how to obtain normalized probabilities of observations from the quantum state of the universe.  This is particularly a problem when eternal inflation leads to a universe of unbounded size so that there are apparently infinitely many realizations or occurrences of observations of each of many different kinds or types, making the ratios ambiguous.  There is also the danger of domination by Boltzmann Brains.  Here two new Spacetime Average Density (SAD) measures are proposed, Maximal Average Density (MAD) and Biased Average Density (BAD), for getting a finite number of observation occurrences by using properties of the Spacetime Average Density (SAD) of observation occurrences to restrict to finite regions of spacetimes that have a preferred beginning or bounce hypersurface.  These measures avoid Boltzmann brain domination and appear to give results consistent with other observations that are problematic for other widely used measures, such as the observation of a positive cosmological constant.

\end{abstract}
\normalsize
\baselineskip 18.8 pt
\newpage
\section{Introduction}

Theoretical cosmology is plagued with the {\it measure problem}, the problem of how to predict the probabilities of observations in the universe from the quantum state (or from some other such input).  
(See \cite{Freivogel:2011eg} for a fairly recent review, and see the 2014 June 2 Version 1 of this paper on the arXiv for a list of 245 references that I had found then that seemed to be related to the measure problem, though it was viewed as editorially inadvisable to include this excessively long list in the published version of this paper.)  This problem is particularly severe for universes that contain infinitely many observations of each of more than one kind, for then one has the ambiguity of taking ratios of infinite numbers.  However, there is also the challenge even for large finite universes, because of the failure of Born's rule \cite{Page:2008ns, Page:2009qe, Page:2009mb, Page:2010bj}.

A solution to the measure problem should not only be able to give normalized probabilities for observations (that is, normalized so that the sum over all possible observations is unity, not just the sum over all the possible observations of one observer at one time) but also make the normalized probabilities of our observations not too small.  One can take the normalized probability of one's observation as given by some theory as the likelihood of that theory.  (Here a theory includes not only the quantum state but also the rules for getting the probabilities of observations from it.)  Then if one weights the likelihoods of different theories by the prior probabilities one assigns to them and normalizes the resulting product, one gets the posterior probabilities of the theories.  One would like to find theories, including their solutions to the measure problem, that give high posterior probabilities.

A major threat to getting high posterior probabilities for theories is the possibility that they may predict that observations are dominated by those of Boltzmann brains that arise from thermal and/or vacuum fluctuations 
\cite{Dyson:2002pf, Albrecht:2002uz, Albrecht:2004ke, Page:2005ur, Yurov:2005xb, Page:2006dt, Bousso:2006xc, Page:2006hr, Linde:2006nw, Page:2006ys, Vilenkin:2006qg, Page:2006nt, Vanchurin:2006xp, Banks:2007ei, Carlip:2007id, Hartle:2007zv, Giddings:2007nu, Giddings:2007pj, Page:2007bt, Li:2007dh, Page:2007ei, Bousso:2007nd, ArkaniHamed:2008ym, Gott:2008ii, Page:2008mx, Freivogel:2008wm, NYTimes, Davenport:2010jy, Page:2010re, Boddy:2013qma, Boddy:2014eba}.  
Most such Boltzmann brain observations seem likely to be much more disordered than ours, so if the probabilities of our ordered observations are diluted by Boltzmann brain observations in theories in which they dominate the probabilities, that would greatly reduce the likelihood of such theories.  Therefore, one seeks theories with solutions to the measure problem that suppress Boltzmann brains if this can be done without too great a cost in complexity that would tend to suppress the prior probabilities assigned to such theories.

Most proposed solutions to the measure problem of cosmology tend to lean toward one or the other of two extremes.  Some, particularly those proposed by Hartle, Hawking, Hertog, and/or Srednicki \cite{Hartle:1997up, Hawking:2002af, Hartle:2002nq, Hartle:2004qv, Hartle:2007zv, Hawking:2007vf, Hartle:2007gi, Hartle:2008fw, Hartle:2008ng, Hartle:2009ig, Srednicki:2009vb, Hartle:2010vi, Srednicki:2010yw, Hartle:2010dq, Hertog:2011ky, Hartle:2012tv, Hartle:2013oda}, which often apply the consistent histories or decohering histories formalism 
\cite{Griffiths:1984rx, Omnes:1988ek, Omnes:1988em, Omnes:1988ej, Omnes:1992ag, GellMann:1991av, GellMann:1992kh, GellMann:1989nu, GellMann:1991ck, GellMann:1994wm, GellMann:1995cu, GellMann:2006uj, GellMann:2011dt, Gell-Mann:2013hza} 
to the Hartle-Hawking no-boundary proposal for the quantum state of the universe \cite{Hartle:1983ai}, tend to suggest that the measure is determined nearly uniquely by the quantum state (at least if a fairly-unique typicality assumption is made \cite{Hartle:2007zv, Srednicki:2009vb, Srednicki:2010yw}).  Others, particularly those proposed in the large fraction of papers cited in Version 1 of this paper for the measure problem that are focused on eternal inflation, tend to suggest that the quantum state is mostly irrelevant and that the results depend mainly on how the measure is chosen in the asymptotic future of an eternally inflating spacetime.  Here, motivated by considerations I have expressed previously on Born's rule \cite{Page:2008ns, Page:2009qe, Page:2009mb, Page:2010bj}, on connecting observations to the quantum state \cite{Page:1994gd, Page:1995dc, Page:1995nj, Page:1995kw, Page:1994pp, Page:2001ba, Page:2006er, Page:2011sr}, and on whether observational probabilities can be independent of the quantum state \cite{Page:2009dp}, I shall steer a middle course and suggest that both the quantum state and the measure are crucial, further suggesting that the measure is dominated by observations not too late in the spacetime.

If the gross asymptotic behavior of an eternally inflating universe is insensitive to the details of the quantum state (say other than requiring that the state be within some open set), then I would think it implausible that the relative probabilities of observations would depend only on the gross asymptotic behavior of the universe.  Therefore, I do not favor measures that have temporal cutoffs that are eventually taken to infinity {\it and} have the property that for a very large finite cutoff they depend mainly on the properties of the spacetime at very large times (e.g., times near the cutoff).  If there is a time-dependent weighting to the measure, I suspect that it should not depend mainly on the asymptotic behavior.  In particular, I am sceptical of the specific form of ``{\it Assumption 3.  Typicality.}'' of Freivogel \cite{Freivogel:2011eg} that ``we are equally likely to be anywhere consistent with our data \ldots [so that with] a finite probability for eternal inflation, which results in an infinite number of observations, \ldots we can ignore any finite number of observations.''  In a footnote to this statement, Freivogel admits, ``This conclusion relies on an assumption about how to implement the typicality assumption when there is a probability distribution over how many observations occur \cite{Page:2010bj}.''  In this paper I shall reject this assumption, which Freivogel notes \cite{Freivogel:2011eg} that I have called ``{\it observational averaging}'' \cite{Page:2010bj}, and instead investigate non-uniform measures over spacetimes that suppress the asymptotic behavior.

As an example of what I mean, in \cite{Page:2008zh} (see also \cite{Page:2008ns, Page:2009qe, Page:2009mb, Page:2011gq} for further discussion and motivation) I proposed volume averaging instead of volume weighting to avoid divergences in the measure of Boltzmann brain observations on spatial hypersurfaces as they expand to become infinitely large.  However, summing up over all hypersurfaces still gave a divergence if that were done by a uniform integral over proper time $t$ and if indeed the proper time goes to infinity.  One could make this integral finite by cutting it off at some finite upper bound to the proper time, say $t_*$, but then as $t_*$ is taken to infinity, asymptotically half of the integral would be given by times within a factor of two of the temporal cutoff $t_*$.  Therefore, as $t_*$ is taken to infinity, the relative probabilities will be determined by the asymptotic behavior of the spacetime.  For example, if it were an asymptotically de Sitter spacetime that does not have bubble nucleation to new hot big bang regions that lead to a sufficiently large number of ordinary observers, the relative probabilities will apparently be dominated by Boltzmann brains in the asymptotic de Sitter spacetime.  Even if de Sitter keeps nucleating new big bangs at a sufficient rate for ordinary observers produced by these big bangs to dominate over Boltzmann brains in the expanding regions that remain asymptotically de Sitter, if a transition occurs to Minkowski spacetime that cannot nucleate new big bangs, and if Boltzmann brains can indeed form in the vacuum state in Minkowski spacetime
\cite{Page:2005ur, Page:2006dt, Page:2006hr, Page:2006ys, Page:2006nt, Page:2007bt, Page:2007ei, Page:2008mx, Davenport:2010jy, Page:2010re, Boddy:2014eba}, 
they will eventually dominate over ordinary observers if the weighting is uniform over proper time up to some cutoff $t_*$ that is taken to infinity.

Therefore, in \cite{Page:2010re} I proposed Agnesi weighting, integrating over $dt/(1+t^2)$ (with the proper time $t$ measured in Planck units) rather than over $dt$, the uniform integral over proper time.  In this case the measure will be dominated by finite times even without a cutoff.  Alternatively, if one did continue to use a cutoff $t_*$, the range of times which dominates the integral will not grow indefinitely as $t_*$ is taken to infinity but instead will remain at fixed finite times (assuming a measure on hypersurfaces that does not diverge as the hypersurfaces become larger and larger).

When Agnesi weighting \cite{Page:2010re} is combined with volume averaging \cite{Page:2008zh, Page:2011gq}, it appears to be statistically consistent with all observations and seems to give much higher likelihoods than measures using the approaches of Hartle, Hawking, Hertog, and/or Srednicki.   It does not require the unproven hypothesis that bubble nucleation rates for new big bangs are higher than Boltzmann brain nucleation rates \cite{Bousso:2008hz, Bousso:2006xc, DeSimone:2008if, Freivogel:2008wm, Freivogel:2011eg}, as the most popular eternal inflation measures require \cite{Freivogel:2011eg}.  It also does not lead to measures dominated by observations of a negative cosmological constant \cite{Salem:2009eh, Bousso:2009gx, Freivogel:2011eg}, which is contrary to what our observations give.  Therefore, for fitting observations without needing to invoke unproven hypotheses, it seems to be the best measure proposed so far.

On the other hand, Agnesi weighting is admittedly quite {\it ad hoc}, so there is no obvious reason why it should be right.  Ideally one would like to find a measure that is more compellingly elegant and simple and which also gives high likelihoods for theories using it and also having elegant and simple quantum states.  However, since none of us have found such a measure, it may be worthwhile to investigate other alternatives to Agnesi weighting.

\section{Spacetime Average Density (SAD) Measures}

Here I wish to propose new solutions to the measure problem with a weighted distribution over variable, rather than fixed, proper-time cutoffs depending on the spacetime average density of observation occurrences up to the cutoff.  The {\it ad hoc} weighting function $dt/(1+t^2)$ of Agnesi weighting will be eliminated, though at the cost of two different rather {\it ad hoc} algorithms for constructing a weighting over proper time to damp the late-time contribution to the measure for observations.

Let me use the index $i$ to denote the theory $T_i$, which I take to include not only the quantum state of the universe but also the rules for getting the probabilities of the observations from the quantum state.  I shall assume that the quantum state given by $i$ gives, as the expectation value of some positive operator depending upon the spacetime and perhaps also upon the theory, a relative probability distribution or measure $\mu_{ij}$ for different quasiclassical inextendible spacetimes $S_j$, labeled by the index $j$, that each has definite occurrences of the observation $O_k$, labeled by the index $k$, occurring within the spacetime at definite location regions that I shall assume are much smaller than the spacetime itself and so can be idealized to be at points within the spacetime.  Because each spacetime includes definite observation occurrences at definite locations, it is not simply a manifold with a metric (though it needs to have that as well) but is a spacetime description of all the observation occurrences within it.  

I should note that $k$ labels the complete content of the observation $O_k$ (which is generically not sufficient to specify uniquely the location of an occurrence of the observation $O_k$ within a spacetime $S_j$), so that there can be multiple occurrences of the observation $O_k$ at different locations within the spacetime.  However, since I am assuming that different observations as such are intrinsically distinguished only by their content and not by their locations, the probability of an observation $O_k$ given by a theory $T_i$,
\begin{equation}
P_{ik} = P(O_k|T_i),
\label{observation-probability}
\end{equation}
is the total normalized measure for all occurrences of the observation $O_k$ at all the different locations at which the observation occurs in each spacetime $S_j$ and in all spacetimes $S_j$ given by the quantum state specified by the theory $T_i$ in which the observation occurs.  Besides specifying the quantum state, a theory $T_i$ must also specify how to get the total measure for the observation $O_k$ from the quantum state.  In the present proposals, I am assuming that this measure is obtained by a suitably weighted sum (depending on the spacetime $S_j$ that specifies not only a geometry but also the locations and types of all observation occurrences within it) of the occurrences of the observation $O_k$ within a spacetime $S_j$, further weighted by the measure $\mu_{ij}$ for the spacetime $S_j$ in the quantum state given by the theory $T_i$, and then summed over all spacetimes $S_j$.  I shall also assume that the measure $\mu_{ij}$ is a linear functional of the quantum state, given by the expectation value in the quantum state of a suitable positive operator for the existence of that spacetime.  Then the probability $P_{ik}$ of the observation $O_k$ given the theory $T_i$ would be the normalized expectation value of some positive operator (depending upon the operators for the existence of the spacetimes $S_j$ and upon the weighting for the observation $O_k$ that may occur multiple times within various ones of these spacetimes), as proposed in my formalism for what I have called Sensible Quantum Mechanics \cite{Page:1995dc, Page:1995nj, Page:1995kw, Page:1994pp, Page:2006er} or Mindless Sensationalism \cite{Page:2001ba, Page:2011sr}.

I shall also assume that each such spacetime $S_j$ with positive measure $\mu_{ij}$ in the theories $T_i$ that I shall be considering is globally hyperbolic with compact Cauchy surfaces and has the equivalent of an initial compact Cauchy hypersurface, either the Terminally Indecomposable Future Set (TIFS) \cite{Geroch:1972un, Hawking:1973uf} of a big bang, or the extremal hypersurface of globally minimum volume for spacetime with a bounce.  Then I shall introduce a time function $t$ that is the supremum of the absolute value of the proper time of any causal curve from the point where it is evaluated to this preferred initial hypersurface.  Next, evaluate the 4-volume $V_j(t)$ of the spacetime region, say $R_{jt}$, in the spacetime $S_j$ that is within a time $t$ or less of the preferred initial hypersurface, and the number $N_{jk}(t)$ of occurrences of the observation $O_k$ that occur within this spacetime region $R_{jt}$ that are of type $k$.  The sum of the number of occurrences of all observation types $O_k$ that occur within this spacetime region of spacetime $S_j$ up to time $t$ is
\begin{equation}
N_j(t) = \sum_k N_{jk}(t).
\label{number-of-observations}
\end{equation}

A subtlety is the fact that presumably different types of observations $O_k$ have different measures as well as different spacetime frequencies of occurring.  Therefore, $N_{jk}$ should not literally be the number of occurrences of the observation $O_k$, but some sort of weighted number, weighted by some factor that depends upon the particular observation type $O_k$ 
\cite{Page:1995dc, Page:1995nj, Page:1995kw, Page:1994pp, Page:2001ba, Carter:2003yj, Page:2006er, Page:2011sr, Carter:2012mh}.  
For example, one might suppose that on earth there are far more ant observations than human observations, but it seems plausible that most human observations have much greater weight than most ant observations (perhaps correlated with the generally increased complexity of human observations over ant observations, though I do not know of any unique obvious detailed form for this correlation).  Then the weighted number $N_{jk}$ of human observations could be greater than that of ant observations, despite the greater number of ant observations if they were simply counted equally weighted, helping to make it not statistically surprising why we find ourselves experiencing human observations rather than ant observations.  Henceforth I shall assume that the numbers of occurrences $N_{jk}$ in the spacetime $S_j$ of the observation $O_k$, and the total number of occurrences in that spacetime of all observations, are weighted numbers of occurrences.

From these quantities, calculate the Spacetime Average Density of occurrences of the observation $O_k$ of type $k$ for the spacetime region $R_{jt}$ of 4-volume $V_j(t)$ within time $t$ of the preferred initial hypersurface in the spacetime $S_j$,
\begin{equation}
\bar{n}_{jk}(t) = N_{jk}(t)/V_j(t).
\label{SADk}
\end{equation}
The sum of these over all observation types $O_k$ is the total Spacetime Average Density of all observation occurrences in the spacetime region $R_{jt}$,
\begin{equation}
\bar{n}_j(t) = \sum_k \bar{n}_{jk}(t) = N_j(t)/V_j(t).
\label{SAD}
\end{equation}
I shall call this the SAD of that region, or the SAD function of $t$ for the spacetime $S_j$.

Now I wish to construct spacetime analogues of spatial volume averaging \cite{Page:2008zh}, weighting each observation $O_k$ by some form of its Spacetime Average Density.  The simplest procedure would appear to be to take the full Spacetime Average Density of each observation $O_k$ over all of each spacetime $S_j$ and then weight by the quantum measures $\mu_{ij}$ that the theory $T_i$ assigns to that spacetime.  However, if the spacetime extends to arbitrarily large $t$ and asymptotes in some region that locally approaches the vacuum with a positive density per 4-volume of Boltzmann brains, the full Spacetime Average Density will be dominated by Boltzmann brain observations that are presumably nearly all highly disordered.  Then the normalized probabilities of ordered observations such as our own would be very low (having been diluted by the enormous number of disordered observations $O_k$ that each have similar measures if made by Boltzmann brains), giving a very low likelihood for such a theory.

One might suppose that if the quantum measure $\mu_{ij}$ in theory $T_i$ for spacetimes $S_j$ with finite total ages (finite bounds on the proper time $t$) is not so small, relative to the measure of spacetimes that last forever, as the ratio of the Spacetime Average Density of ordinary observers in the finite-age spacetimes to the Spacetime Average Density of Boltzmann brains in the infinite-age spacetimes, that then when weighted by the Spacetime Average Density of all observations, the finite-age ones will dominate, giving a cosmic doomsday argument for a spacetime with a finite total age, as indeed I have previously proposed \cite{Page:2009mc}.  However, like many of the current inflationary universe measures \cite{Salem:2009eh, Bousso:2009gx, Freivogel:2011eg}, this appears to be dominated by observations of a negative cosmological constant, contrary to what we see, so in this paper I shall seek other measures that avoid this problem.

Therefore, I shall seek a weighting over different spacetime regions that avoids domination by both Boltzmann brains and by a negative cosmological constant.  The Spacetime Average Density over spacetime regions $R_{jt}$ for certain reasonable values of $t$ that are of the same order as what is believed to be the present age of our universe, about 14 billion years, would seem to be dominated by ordinary observations and not by Boltzmann brains, at least for a suitable quantum state of the universe (though perhaps not for the Hartle-Hawking no-boundary proposal \cite{Hartle:1983ai}, which seems to predict mostly nearly empty de Sitter spacetime that would apparently be dominated by Boltzmann brains even for times much shorter than the time needed for Boltzmann brains to dominate in a universe that starts with a hot big bang \cite{Page:2006hr}).  However, I do not want to introduce some fixed parameter value for what $t$ is for the spacetime regions $R_{jt}$ to be used for the Spacetime Average Densities of the various observations.

Instead, I shall seek a measure to be given in terms of an auxiliary function $f_{ij}(t)$, determined both by the theory $T_i$ and by the spacetime $S_j$ existing within the theory with quantum measure $\mu_{ij}$, which increases monotonically from 0 to 1 as $t$ ranges from 0 to $\infty$ within the spacetime $S_j$.  (If $t$ runs only from 0 to $t_j < \infty$ for some spacetimes $S_j$, I shall require that $f_{ij}(t)$ increase monotonically from 0 to 1 as $t$ increases from 0 to $t_j$ and then stay at 1 for all values of $t$ greater than $t_j$ that do not actually occur within the spacetime $S_j$, so that for simplicity I can take $t$ running from 0 to $\infty$ for each spacetime.)  I shall then assume that equal ranges of $f_{ij}(t)$ contribute equally to the measure in choosing the value of $t$ used to cutoff the spacetime.  

First I shall explain more explicitly how to use $f_{ij}(t)$ to get the measure, and then I shall postulate different ways (labeled by the index $i$ in the theory $T_i$ that includes not only the quantum measures $\mu_{ij}$ for the different spacetimes $S_j$ but also the rules for getting the measure for converting the quantum state to observational probabilities) to get the auxiliary function $f_{ij}(t)$ from the SAD function $\bar{n}_j(t)$ for the spacetime $S_j$.  In particular, I shall propose that the weighted Spacetime Average Density for the occurrences of the observation $O_k$ in theory $T_i$ and in spacetime $S_j$ with auxiliary function $f_{ij}(t)$ is
\begin{equation}
\bar{n}_{ijk} = \int_0^\infty dt \frac{df_{ij}}{dt} \bar{n}_{jk}(t)
= \int_0^\infty dt \frac{df_{ij}}{dt} \frac{N_{jk}(t)}{V_j(t)}.
\label{3-index-weighted-average-density}
\end{equation}
The weighted Spacetime Average Density in theory $T_i$ and spacetime $S_j$ for all observations $O_k$ is the sum of this over the $k$ that labels the observations:
\begin{equation}
\bar{n}_{ij} = \sum_k \bar{n}_{ijk} = \int_0^\infty dt \frac{df_{ij}}{dt} \bar{n}_{j}(t)
= \int_0^\infty dt \frac{df_{ij}}{dt} \frac{N_j(t)}{V_j(t)}.
\label{2-index-weighted-average-density}
\end{equation}

Next, to include the quantum measure $\mu_{ij}$ that the theory $T_i$ assigns to the spacetime $S_k$, I propose that the unnormalized measure or relative probability $p_{ik}$ in theory $T_i$ of the observation $O_k$ is the sum of the weighted Spacetime Average Densities further weighted by the quantum measures:
\begin{equation}
p_{ik} = \sum_j \mu_{ij}\bar{n}_{ijk}.
\label{relative-probabilities}
\end{equation}
Finally, dividing by the normalization factor
\begin{equation}
p_i = \sum_k p_{ik} = \sum_j \mu_{ij}\bar{n}_{ij}
\label{sum-of-relative-probabilities}
\end{equation}
gives the normalized probability in the theory $T_i$ of the observation $O_k$ as
\begin{equation}
P_{ik} \equiv P(O_k|T_i) = \frac{p_{ik}}{p_i}
= \frac{\sum_j \mu_{ij} \bar{n}_{ijk}}{\sum_{j,k} \mu_{ij} \bar{n}_{ijk}}.
\label{normalized-probabilities}
\end{equation}
Of course, it remains to be said what different theories $T_i$ give for the way to get the auxiliary function $f_{ij}(t)$ from the SAD function $\bar{n}_j(t)$ for the spacetime $S_j$.

\subsection{Maximal Average Density (MAD) Measure}

First, consider theories $T_i$ that employ what I shall call the Maximal Average Density (MAD) measure.  These make use of the time $t_{*j}$ that is the value of $t$ that gives the global maximum value of the SAD function of $t$ for the spacetime $S_j$, $\bar{n}_j(t) = N_j(t)/V_j(t)$, the Spacetime Average Density of the total occurrences of all observations up to proper time $t$ in the spacetime $S_j$.  That is, $\bar{n}_j(t) \leq \bar{n}_j(t_{*j})$ for all $t$ in the spacetime $S_j$.  (For simplicity, I shall assume that there is zero quantum measure $\mu_{ij}$ for spacetimes with more than one value of $t_{*j}$ at which $\bar{n}_j(t)$ attains its global maximum, so that $\bar{n}_j(t) < \bar{n}_j(t_{*j})$ for all $t \neq t_{*j}$ in all spacetimes $S_j$ with positive measures.)

In particular, the Maximal Average Density or MAD measure is the one in which
\begin{equation}
f_{ij}(t) = \theta(t-t_{*j}),
\label{MAD-f}
\end{equation}
the Heaviside step function, being 0 for times $t$ before the global maximum for $\bar{n}_j(t)$ and being 1 for times after this global maximum.  Then $df_{ij}/dt = \delta(t-t_{*j})$, a Dirac delta function centered on the global maximum for $\bar{n}_j(t)$ for the spacetime $S_j$, so Eq.\ (\ref{3-index-weighted-average-density}) gives
\begin{equation}
\bar{n}_{ijk} = \bar{n}_{jk}(t_{*j}).
\label{MAD-3-index-weighted-average-density}
\end{equation}
This then leads to the normalized probability for the observation $O_k$ given a MAD theory $T_i$ as being
\begin{equation}
P_{ik} \equiv P(O_k|T_i) 
= \frac{\sum_j \mu_{ij} \bar{n}_{jk}(t_{*j})} {\sum_{j,k}\mu_{ij}\bar{n}_{jk}(t_{*j})}.
\label{MAD-normalized-probabilities}
\end{equation}
Of course, there is not a unique MAD theory, since for this MAD function $f_{ij}(t) = \theta(t-t_{*j})$, there are many different MAD theories giving different quantum states and hence different quantum measures $\mu_{ij}$ for the spacetimes $S_j$.

One might suppose that a typical inextendible spacetime which gives a large contribution to the probability $P_{ik}$ in a plausible theory $T_i$ of a typical human observation $O_k$ would have something like a big bang at small $t$ (though perhaps actually a bounce at $t=0$ \cite{Page:2009ct}), a relatively low density of observation occurrences until a period around $t \sim t_0$ when planets heated by stars exist and have a relatively high density of observation occurrences produced by life on the warm planets (compared with that at any greatly different time), and then a density of observation occurrences that drops drastically as stars burn out and planets freeze, until the density of observation occurrences asymptotes to some very tiny but still positive spacetime density of Boltzmann brain observations.  In this case it is plausible to expect that $\bar{n}_{ij}(t)$ will start very small for small $t$ when the spacetime $S_j$ is too hot for life (and when life has not had much time to evolve), rise to a maximum at a time $t \sim t_0$ when planetary life prevails, and then drops to a very small positive asymptotic constant when planetary life dies out and Boltzmann brains dominate.

For example, a mnemonic $k=0$ $\Lambda$CDM Friedmann-Lema\^{\i}tre-Robertson-Walker model \cite{Scott:2013oib} with $\Lambda = 3 H^2_\infty \approx (10\ \mathrm{Gyr})^{-2} \approx \mathit{ten\ square\ attohertz} \approx 3\pi/(5^3 2^{400})$ Planck units (and so fairly accurately applicable to our universe only after the end of radiation dominance but here used for all times) gives
\begin{equation}
V_j(t) \approx V_3\, (2/27)\, H_\infty^{-1}\, x(t) \equiv 
V_3\, (2/27)\, H_\infty^{-1}\, [\sinh{(3 H_\infty t)} - 3 H_\infty t]
\label{4-volume}
\end{equation}
with $V_3$ the present 3-volume (unknown and perhaps very large because the universe appears to extend far beyond what we can see of it, though here I shall assume that it is finite) and $x(t) = \sinh{y(t)} - y(t)$ with $y(t) = 3 H_\infty t = \sqrt{3\Lambda}\, t$.

A very crude toy model for the SAD function $\bar{n}_j(t)$ for the Spacetime Average Density of all observations might be
\begin{equation}
\bar{n}_j(t) \sim A \left(\frac{x(t)}{1+x(t)^2} + \epsilon\right),
\label{SAD-density}
\end{equation}
where $A$ is an unknown constant that parametrizes the peak density of ordinary observations that are represented by the first term that rises and then falls, and where $A\, \epsilon$ is the much, much smaller density of Boltzmann brain observations that is crudely assumed to be constant.  (For a Boltzmann brain that is the vacuum fluctuation of a human-sized brain, one might expect $\epsilon \sim 10^{-10^{42}}$ \cite{Page:2006hr}.)  The total number of ordinary observation occurrences in this crude model is finite, $A\, V_3$, but the total number of Boltzmann brain observations grows linearly with the 4-volume $V_j(t) = V_3\, x(t)$ and hence diverges if indeed $t$ and $x(t)$ go to infinity as assumed.  The SAD function rises from $A\, \epsilon$ at $t = 0$ and $x = 0$ (probably an overestimate, as I would suspect that when the universe is extremely dense, Boltzmann brain production would be suppressed, but since $\epsilon \ll 1$ I shall ignore this tiny error, no doubt much smaller than the error of the crude time-dependent term for the Spacetime Average Density of ordinary observations) monotonically to $A(0.5 + \epsilon)$ at $t = t_{*j}$ that gives $x_{*j}\equiv x(t_{*j}) = 1$ and then drops monotonically back to $A\, \epsilon$ at $t = \infty$ that gives $x = \infty$.

Then the MAD measure gives
\begin{equation}
\bar{n}_{ij} = \bar{n}_{j}(t_{*j}) = A(0.5 + \epsilon).
\label{MAD-2-index-weighted-average-density}
\end{equation}
The first term in the sum corresponds to ordinary observations, and the second corresponds to Boltzmann brain observations.  If all the different spacetimes $S_j$ that have positive quantum measure $\mu_{ij}$ had SAD functions $\bar{n}_j(t)$ that were proportional to this one (with the same ratio of ordinary and Boltzmann brain observations), then the total normalized probability for Boltzmann brain observations would be only $\epsilon/(0.5 + \epsilon) \approx 2 \epsilon \ll 1$.  Thus the MAD measure would solve the Boltzmann brain problem, even for models in which Boltzmann brain production is faster than the production of new bubble universes and even for models that asymptote to Minkowski spacetime with its infinite spacetime volume and presumed positive density per 4-volume of Boltzmann brain observations that would cause Boltzmann brain domination in most other proposed solutions to the measure problem.

\newpage

\subsection{Biased Average Density (BAD) Measure}

Next, consider what I shall call the Biased Average Density (BAD) measure.  A motivation for going from MAD to BAD is that the MAD measure does not give any weight to observations within a spacetime that is after the time $t_{*j}$ at which the SAD, $\bar{n}_j(t)$, is maximized.  It does not seem very plausible that any observation occurrence within a spacetime of positive measure would contribute zero weight to that kind of observation $O_k$, so the BAD measure replaces the MAD measure by a weighting that is positive for all observation occurrences within an inextendible spacetime (except possibly for a set of measure zero).  The auxiliary function in the BAD measure is given by
\begin{equation}
f_{ij}(t) = \frac{\int_0^t dt' |d\bar{n}_j(t')/dt'|}
            {\int_0^\infty dt' |d\bar{n}_j(t')/dt'|},
\label{BAD-f}
\end{equation}
which again increases monotonically from 0 at $t = 0$ to 1 at $t = \infty$, but now continuously rather than suddenly jumping from 0 to 1 as the MAD auxiliary function $f_{ij}(t) = \theta (t - t_{*j})$ does.

In particular, if $\bar{n}_j(t)$ increases monotonically from $n_0$ at $t = 0$ to a single local maximum (the global maximum) value $n_{*}$ at $t = t_{*j}$ and then decreases monotonically to $n_\infty$ at $t = \infty$ (where for now I am suppressing the overbar and $j$ index on $\bar{n}_j(t)$ at these three special times), then
\begin{equation}
f_{ij}(t) = 
  \theta(t_{*j} - t) \frac{\bar{n}_j(t) - n_0}{2 n_{*} - n_0 - n_\infty}
+\theta(t-t_{*j})\frac{2n_{*}-n_0-\bar{n}_j(t)}{2n_{*}-n_0-n_\infty},
\label{BAD-f-one-peak}
\end{equation}
Then Eq.\ (\ref{2-index-weighted-average-density}) gives
\begin{equation}
\bar{n}_{ij} = \frac{2n_{*}^2 - n_0^2 - n_\infty^2}
{2(2n_{*} - n_0 - n_\infty)}.
\label{BAD-2-index-weighted-average-density}
\end{equation}

In the case in which $n_0 = n_\infty$ (as was assumed above in the crude toy model), one gets
\begin{equation}
\bar{n}_{ij} = \frac{1}{2}(n_{*} + n_0).
\label{BAD-special-2-index-weighted-average-density}
\end{equation}
This is what was assumed above in the crude toy model, which has $n_0 = n_\infty = A\, \epsilon$ and $n_{*} = A(0.5 + \epsilon)$, giving
\begin{equation}
\bar{n}_{ij} = A(0.25 + \epsilon).
\label{BAD-special-crude-2-index-weighted-average-density}
\end{equation}

Again taking the first term in the sum to correspond to ordinary observations and the second term to correspond to Boltzmann brain observations, and assuming that all the different spacetimes with positive quantum measure given the same ratio of the first term to the second term, one gets that the total normalized probability for Boltzmann brain observations in the BAD measure would be $\epsilon/(0.25 + \epsilon) \approx 4 \epsilon \ll 1$, roughly twice what it would be in the MAD measure but still extremely small.  Therefore, both the MAD and BAD measures would solve the Boltzmann brain problem.

\section{Conclusions}

\baselineskip 18.1 pt

One might compare the MAD and BAD measures, which are SAD measures using the Spacetime Average Density, with the Agnesi measure \cite{Page:2010re}, which uses spatial averaging over hypersurfaces and a weighting of hypersurfaces by the Agnesi function of time, $dt/(1+t^2)$ with time $t$ in Planck units.  In some ways the MAD and BAD measures appear to have more complicated algorithms, but they do avoid the use of an explicit {\it ad hoc} function of time such as the Agnesi function, though it is one of the simplest functions that is positive and gives a finite integral over the real axis.

The weighting factor of the Agnesi measure does favor earlier times or youngness (as both the MAD and BAD measures do in different ways), but in a fairly weak or light way, without exponential damping in time.  Therefore, it might be called a Utility Giving Light Youngness (UGLY) measure.  As a result, I have now made alternative proposals for measures that are MAD, BAD, and UGLY.  I am still looking for one that is GOOD in a supreme way of giving a high posterior probability by both giving a likelihood (probability of one's observation given the theory that includes the measure) that is not too low (which it seems that all three of my proposed measures would do with a suitable quantum state, such as perhaps the Symmetric Bounce state \cite{Page:2009ct}) and giving a prior probability (assumed to be higher for simpler or more elegant theories) that is not too low (which my measures might not be in comparison with a yet unknown measure that one might hope could be much simpler and more elegant).  However, if GOOD were interpreted as simply meaning Great Ordinary Observer Dominance, then since all three of my proposed measures suppress Boltzmann brains relative to ordinary observers, one could say that the MAD, the BAD, and the UGLY are all GOOD.

\section{Acknowledgments}

I have benefited from discussions with many colleagues on the measure problem in cosmology, of which a proper subset that presently comes to mind includes Andy Albrecht, Tom Banks, Andrei Barvinsky, Raphael Bousso, Adam Brown, Steven Carlip, Sean Carroll, Brandon Carter, Willy Fischler, Ben Freivogel, Gary Gibbons, Steve Giddings, Daniel Harlow, Jim Hartle, Stephen Hawking, Simeon Hellerman, Thomas Hertog, Gary Horowitz, Ted Jacobson, Shamit Kachru, Matt Kleban, Stefan Leichenauer, Juan Maldacena, Don Marolf, Yasunori Nomura, Joe Polchinski, Steve Shenker, Eva Silverstein, Mark Srednicki, Rafael Sorkin, Douglas Stanford, Andy Strominger, Lenny Susskind, Bill Unruh, Erik Verlinde, Herman Verlinde, Aron Wall, Nick Warner, and Edward Witten.  Face-to-face conversations with Gary Gibbons, Jim Hartle, Stephen Hawking, Thomas Hertog, and others were enabled by the gracious hospitality of the Mitchell family and Texas A \& M University at a workshop at Great Brampton House, Herefordshire, England.  This work was supported in part by the Natural Sciences and Engineering Research Council of Canada.

\end{document}